\documentclass{appolb}
\usepackage{epsfig}
\usepackage[latin1]{inputenc}
\usepackage{xspace}
\usepackage{cite}
\newcommand{\AROMA}{A\scalebox{0.8}{ROMA}\xspace}
\newcommand{\GRAPE}{G\scalebox{0.8}{RAPE}\xspace}
\newcommand{\jetrad}{J\scalebox{0.8}{ETRAD}\xspace}
\newcommand{\dispatch}{D\scalebox{0.8}{ISPATCH}\xspace}
\newcommand{\disent}{D\scalebox{0.8}{ISENT}\xspace}
\newcommand{\disaster}{D\scalebox{0.8}{ISASTER++}\xspace}
\newcommand{\apacic}{\scalebox{0.8}{APACIC++}\xspace}
\newcommand{\herwig}{\scalebox{0.8}{HERWIG}\xspace}

\newcommand{\diclus}{D\scalebox{0.8}{ICLUS}\xspace}
\newcommand{\pythia}{P\scalebox{0.8}{YTHIA}\xspace}
\newcommand{\lepto}{L\scalebox{0.8}{EPTO}\xspace}

\newcommand{\rapgap}{R\scalebox{0.8}{APGAP}\xspace}

\newcommand{\ariadne}{A\scalebox{0.8}{RIADNE}\xspace}
\newcommand{\jimmy}{J\scalebox{0.8}{IMMY}\xspace}

\newcommand{\cascade}{C\scalebox{0.8}{ASCADE}\xspace}

\newcommand{\as}{\ensuremath{\alpha_{s}}\xspace}

\newcommand{\particle}[1]{\ensuremath{\mathrm{#1}}}
\newcommand{\antiparticle}[1]{\ensuremath{\bar{\mathrm{#1}}}}

\newcommand{\qs}{\particle{s}}

\newcommand{\sbar}{\antiparticle{s}}

\newcommand{\ee}{\ensuremath{e^+e^-}\xspace}

\newcommand{\lam}{\ensuremath{\Lambda_{\mathrm{QCD}}}\xspace}
\def\mrm#1{\mathrm{#1}}
\def\sub#1{\ensuremath{_{\mrm{#1}}}}

\def\f2d3{\ensuremath{F_2^{\mrm{D}3}}}
\preprint{~}

\usepackage{epsfig}

\begin{document}
\title{Multi-hadron Final States%
\thanks{Presented at X International Workshop on Deep Inelastic 
Scattering (DIS2002), Krak\'ow, Poland, 30 April - 4 May, 2002}%
}
\author{Leif L\"onnblad
\address{Dept. of Theoretical Physics, Lund University, Lund, Sweden}
\\ $~$ \\
Jos\'e Repond
\address{Argonne National Laboratory, Argonne, IL 60439}
\and
Marek Zieli\'nski
\address{Dept. of Physics and Astronomy, University of Rochester, Rochester, NY 14627}
}
\maketitle
\begin{abstract}
This summary aims to highlight major results and insights gained from recent 
studies of hadronic final states in $ep$, $p\overline{p}$, $e^+e^-$, as well as
relevant theoretical developments, presented in the \emph{Multi-hadron 
final states} parallel sessions of the DIS2002 workshop held
in Krak\'ow, Poland in May 2002.
\end{abstract}
  
\section{Introduction}

Presentations at the \emph{Multi-hadron final states} sessions covered
a wealth of recent results on jet production characteristics and jet
properties, heavy flavor production and decays, dimuon
photoproduction, instanton searches, small-$x$ final states, studies
of hadronization, and theoretical progress in resummation and parton
shower formalisms, providing new insights into and sensitivity to a
broad range of physics aspects.  Due to space limitation, we refer the
interested reader to individual papers in these proceedings for exact
definitions of variables, experimental selections, plots, and further
references.

\section{Physics of Jets}

Jet production processes in $ep$, $e^+e^-$ and hadronic interactions continue
to be primary tools for studies of parton dynamics and for testing specific
theoretical descriptions and Monte Carlo models (MC) based on QCD.

While results from HERA on inclusive, dijet, and multi-jet reactions,
induced by real and virtual photons, dominated the discussion, they were
complemented by recent multi-jet measurements at LEP that directly probe
the color structure and the coupling constant of strong interactions, and
by a variety of jet studies at the Tevatron, covering a large range
of jet $E_T$ and probing QCD aspects from the hardest partonic scatter
to the properties of the soft underlying and minimum bias events.

\subsection{Inclusive and dijet results from HERA, plus some photons}

The results presented in the sessions were generally in an impressive 
agreement  with theoretical predictions, for many kinematic variables, and 
for several orders of
magnitude of cross sections. Here, we will attempt to highlight hints of
disagreements, which may point to weaknesses in the description
of particular aspects of the underlying physics. 
Most of the measurements used versions of the 
$k_T$ algorithm and were carried out in
the Breit frame.
In general, the veracity of next-to-leading order (NLO) matrix elements 
is confirmed by good agreement with the
angular distributions in the data (cos$\theta^*$ in dijet 
photoproduction, especially for high dijet mass; 
cos$\theta_3$, $\psi_3$ in 3-jet DIS and in 4-jet photoproduction
at high 4-jet mass.)

The photoproduction data from both H1 and ZEUS probe
the photon  parton distribution functions (PDFs); the
GRV-HO set provides predictions 10-15\% above those from AFG-HO
(and seems to be in better agreement with the data).

\textbf{Jacek Turnau}\cite{DIS02Turnau} presented an H1 measurement of the single inclusive
jet cross section in photoproduction ($Q^2 < 1$ GeV$^2$) 
at high $E_T$ (21--75 GeV), 
where the sensitivity to soft physics is reduced. The data agree 
very well with NLO QCD
and with previous ZEUS results, but because of the large 
uncertainty in the jet energy
scale, the preference for the GRV PDF is only marginal; this uncertainty 
needs to be reduced to take advantage of
the high luminosity measurements at HERA II.

\textbf{Maria Krawczyk}\cite{DIS02Krawczyk} discussed theoretical
issues in the inclusive direct-photon production. There are two
different approaches to counting the powers of $\alpha_s$ in
NLO calculations, leading to different sets of subprocesses
being included; other authors considered the $\log Q^2$ factor present in 
the photon $F_2$ function as an inverse of $\alpha_s$. The ZEUS data
currently have errors too large to discriminate between these
calculations, or between the photon PDFs, but are in rough agreement
with expectations (although none of them describe the data
for rapidities below 0.1).

Dijet photoproduction probes direct and resolved photon components
in more detail and 
is sensitive, already at lowest order, to the gluon content of the photon.
At HERA, the photon structure can be probed at higher scales
than at LEP.  Theoretical calculations are available at NLO.

Dijet photoproduction cross sections from ZEUS, presented by 
\textbf{Anna Lupi}\cite{DIS02Lupi}, are 
sensitive to the photon PDFs, but neither GRV-HO or AFG-HO 
fully describe all 
features of the data. In particular, the predictions do not track the data
when compared as function of the cut on the lower energy jet,
$E_T^{jet2}$. For the cut value used by ZEUS, $E_T^{jet2} > 11$ GeV,
there is a significant difference between data and theory for
$x_\gamma < 0.8$, $x_\gamma$ being the momentum fraction of the photon 
partaking in the hard interaction.  The H1 comparisons, presented by 
\textbf{Gilles Frising}\cite{DIS02Frising},
show a better agreement with theory, which may be related to
the higher cut value ($E_T^{jet2} > 15$ GeV) used by H1.
Frising pointed to significant
NLO scale uncertainty, and sizable hadronization corrections at
high $x_\gamma$. Both speakers concluded
that  HERA data can constrain and should be included in fits to
PDFs;  the constraints could be made more stringent if 
improved higher-order or resummed calculations were available.
The H1 analysis also exhibits some sensitivity to the proton PDFs; 
this could be exploited with the high-statistics HERA II data. 

Up a step in $Q^2$, the region $\Lambda^2_{\rm QCD} << Q^2 < E_T^2$
is the place to investigate contributions from longitudinally
polarized photon interactions, which vanish in the photoproduction 
limit. 
The analysis of H1 data, presented by 
\textbf{Kamil Sedlak}\cite{DIS02Sedlak}, demonstrated 
the importance of the resolved transverse component even in
this $Q^2$ region; however, addition of longitudinal resolved photon
contributions (within the \herwig\ framework) further improves the
agreement with data. Based on a study of $y$-distributions, such
a longitudinal component is preferred to a pure enhancement of the 
transverse part. Interestingly, the predictions of the \cascade\ MC, 
based on the CCFM evolution, describe the data almost as 
well, without explicitly introducing a resolved component (however,
the $Q^2$ dependence at low $x_\gamma$ is poorly described.)

The above analysis neglected the possibility of interference 
between transverse and longitudinal components, which was 
found to be important in some cases studied at LEP. 
\textbf{Urszula Jezuita-D\c{a}browska}\cite{DIS02Jezuita} 
investigated theoretical aspects of
such interference effects in the semi-inclusive Compton process 
in $ep$ collisions. She found that at HERA energies the interference
term is of similar magnitude as the longitudinal term, but of
the opposite sign! The extension of this study to dijet production
at low $Q^2$ would be certainly interesting.

The partonic content of the virtual photon is suppressed  when
$Q^2$ increases, leading to a corresponding suppression of 
the ratio $R$ between cross sections for $x_\gamma < 0.75$
(dominated by the resolved component) and for  $x_\gamma > 0.75$
(dominated by the direct component). 
\textbf{Matthew Lightwood}\cite{DIS02Lightwood} 
investigated this behavior (as seen in the ZEUS 
data) in the case when the dijets were identified as originating
from charm quarks. The charm events were tagged by reconstructing a 
high-$p_T$  $D^{*+} \rightarrow K^-\pi^+\pi^+_s$ decay. After
correcting for the $D^*$ kinematic selections, the $Q^2$
suppression of $R$ for charm events was significantly 
weaker than for the all-flavors case
-- clearly a consequence of the charm-quark mass.
Again, \cascade\ described the charm data well.

A novel approach to the analysis of H1 dijet data in the
DIS regime ($150 < Q^2 < 35000$ GeV$^2$) was discussed 
by \textbf{G\"unter Grindhammer}\cite{DIS02Grindhammer}, 
using a modified Durham algorithm. 
The traditional DIS analyses require
a large inter-jet separation, with only $\approx 10\%$ of
the inclusive sample classified as dijet events.
H1 investigated the minimum separation required for
an adequate agreement with NLO calculation in terms
of the variable $y_2 = {\rm min\;} k^2_{Tij}/scale^2$
(with $scale$ chosen either as $W$ or $Q$).
For $y_2 > 0.001$ the agreement with NLO was excellent,
and the sample retained a third of all DIS events.
Detailed studies of several variables in this expanded
sample showed a very good agreement with NLO calculations,
for two choices of the renormalization scale.
Interestingly, a LO+parton-shower (PS) MC \rapgap\ 
describes the $y_2$-distribution down to
$10^{-5}$, and reproduces other data well, while
\ariadne, \herwig\ and \lepto\ fail at places.

\textbf{Sonja Hillert}\cite{DIS02Hillert} reviewed ZEUS studies 
of internal structure of 
jets in neutral-current, charged-current and photoproduction
data. Such studies provide insights into the transition from
partons to the observed jets of hadrons and are sensitive
to emission of gluons, thus allowing a determination of the strong
coupling constant. The jet substructure is studied through the measurement
of the mean subjet multiplicity; subjets are resolved by reapplying
the $k_T$-clustering algorithm to particles within the jet while using
a smaller resolution parameter $y_{cut}$. 
Using $y_{cut} = 10^{-2}$, and for both NC and CC data, the mean
subjet multiplicity was compared to NLO calculations with the
$\alpha_s$-dependent A-series of CTEQ4 PDFs, yielding 
fits to $\alpha_s$ in good agreement with the PDG value, and
with other ZEUS determinations; the errors are dominated by the
theoretical uncertainty.
Results were also presented for both integrated
and differential energy-density distributions as function of 
the distance from the jet axis. The quark-initiated jet shapes 
determined in DIS events are in good agreement with NLO
expectations and are consistent between CC and NC samples.
In photoproduction, however, one finds a somewhat different
shape, which is expected because of the contributions 
from gluon initiated jets.
A quark-enriched sample has been selected by the $D^*$
charm-tagging technique; these jets have shape distributions
similar to DIS jets.
By taking the fractional contributions of quark and gluon
jets in photoproduction data  from LO MC, and assuming the measured 
shape for the quark jets, ZEUS solved for the shape
of gluon jets and found that \pythia\ provides a good representation.

\subsection{Multi-jet studies at HERA and  LEP}

A direct sensitivity to higher-order effects is obtained through
investigations of multi-jet events. 
Using ZEUS data
\textbf{Claire Gwenlan}\cite{DIS02Gwenlan} presented the first 
four-jet photoproduction
cross section measurement.
Multi-jet production in the photoproduction regime is also
sensitive to the photon structure and to the ensuing multiple parton
interactions.
The distribution of $x_\gamma^{\rm OBS}$ agrees
well with \pythia\  for high mass events, $m_{4\rm J} > 50$ GeV,
but requires a large contribution of multiple parton interactions (MPI)
at lower masses and low $x_\gamma^{\rm OBS}$. Angular distributions
cos$\theta_3$ and $\psi_3$ have been compared to the MC 
expectations. While \herwig\ alone, and \herwig+SUE
(a soft underlying event model) fail, \herwig+ \jimmy\ (a multi-parton
interaction model) work fine, as does \pythia+ MPI;
at high mass the soft effects are much reduced and the LO+PS
generators alone provide a satisfactory description of the data.

In the DIS regime, extra jets are produced primarily through
hard gluon emission, prodding enhanced sensitivity to
$\alpha_s$ and allowing more stringent test of NLO theory.
\textbf{Christian Schwanenberger}\cite{DIS02Schwanenberger} 
presented H1 measurements
of the 3-jet cross section versus several variables, for
$5 < Q^2 < 5000$ GeV$^2$. At lowest $Q^2$ the dominant
theoretical uncertainty is related to the choice of scales, 
but at the high-$Q^2$
end this uncertainty is reduced, thus providing
sensitivity to $\alpha_s$. This sensitivity
is enhanced in the ratio of 2-jet and 3-jet cross sections,
as the scale and some other theoretical and experimental
uncertainties tend to cancel. The data are in good agreement
with NLO (but not LO) using the world average for $\alpha_s$. 
It is notable that the angular cos$\theta_3$ and $\psi_3$
distributions are well described by NLO, but not by phase space,
confirming the expectation for QCD radiation patterns.

Data from $e^+e^-$ annihilation into hadrons provide a
particularly clean environment to study characteristics
of multi-jet events. 
\textbf{Pablo Tortosa}\cite{DIS02Tortosa} presented results from
LEP measurements of 4-jet events. The underlying theoretical
description assumes only a non-abelian gauge symmetry
and standard hadronization models, thus allowing determinations
of the strong coupling constant and the SU(3) color factors
by comparison of the measured 4-jet production rate and
four different angular correlations between jets to NLO 
perturbative predictions. The measurements
are in agreement with SU(3) expectations and with the world 
average value of $\alpha_s(M_Z)$; the uncertainties
on the extracted parameters depend on the number of free parameters
in the fits.
Flavor tagging has been used to enhance the sensitivity to
color factors by identifying quark versus gluon jets.
The results exclude the existence of a massless
gluino at more than 95\% confidence level (it would
modify the factors by adding light fermion degrees 
of freedom).

\subsection{Selected jet results from the Tevatron} 

Traditionally, the Tevatron jet measurements have been
based on the cone algorithm, and were aimed primarily at
the highest transverse energies. In recent years,
however, more attention has been paid to alternative
algorithms and to studies of jet and event structure issues
at low energies.

\textbf{Iain Bertram}\cite{DIS02Bertram} reviewed several 
recent analyses from D\O\
based on the Ellis-Soper version of the $k_T$ algorithm.
D\O\ recently published the inclusive $k_T$ jet cross section,
the first such measurement at the hadron collider.
The measurement includes a study of hadronization
effects, usually neglected for cone jets.
The agreement with NLO theory is not as good as
previously obtained with the cone algorithm --  a behavior that
can be partially traced to energy differences
between jets reconstructed with these algorithms.
The probability of agreement with NLO (\jetrad)
is fair (44\%), but the possibility of some
discrepancy reinvigorated investigations
of issues, such as hadronization, influence of the underlying event,
and properties of the algorithms.
With the $k_T$ algorithm, D\O\ studied 
the jet substructure, similarly to
the discussion above by Hillert. The quark and gluon 
subjet multiplicities have been disentangled
by comparing the measurements at $\sqrt{s}$ =
630 GeV and 1800 GeV and by assuming that
the respective fractions of quark and gluon initiated jets
are known from the MCs. The result for the
ratio between quark and gluon subjet multiplicities
has been found to be in good agreement with \herwig.
A very recent analysis, based on the $k_T$ algorithm,
investigated the event-shape thrust variable $T$, defined here
using only the transverse momenta of the 
two leading jets; despite of this simplification, 
$T$ remains sensitive to extra jets (radiation)
in the event through the imbalance of the
leading jets in the azimuthal plane. The measurement
disagrees with the NLO \jetrad\ description at the very
low values of $1-T$, where resummation effects are
expected to be important,  and at high values of $1-T$,
where the lowest order is $\mathcal{O}(\alpha_s^4)$. Thus,
this measurement provides an excellent opportunity for 
testing future resummed and higher order calculations.
Using the cone algorithm, D\O\ investigated the production
of multi-jet events at low $E_T \approx 20$ GeV
(for inclusive $\geq$1, 2, 3 and  4 jets), compared
to expectations from \pythia\ and \herwig.
Both MCs can be tuned to reproduce the data in
$E_T$ and in various angular distributions, but fail 
to do so with
the default parameters. The sensitive parameter in
\pythia\ is the fraction of core region of hadronic 
matter distribution (PARP(83)), which controls
the rate of multiple parton interactions; in \herwig\
the sensitive parameter is the minimum $p_T$ of the hard
process.

\textbf{Christina Mesropian}\cite{DIS02Mesropian} 
discussed recent analyses from CDF
focused on jet fragmentation properties, jet evolution
from minimum bias events to higher $E_T$, and studies
of properties of the underlying event (UE) in jet
and minimum bias events. These analyses were performed
using variations of the cone algorithm, and utilized both 
calorimeter and tracking information.
Fragmentation into charged particles 
of well-balanced high-mass central dijet events was studied within
the MLLA formalism combined with the LPHD hadronization
prescription. The inclusive momentum distribution of charged
particles in the jet was found to agree well with theoretical
predictions, allowing extraction of the effective momentum
cutoff $Q_{eff}$ at which partons undergo hadronization;
a value of order \lam\ was obtained. The ratio of hadron multiplicities
in quark and gluon jets was also extracted.
For studies of the development and evolution of jets
from low  (0.5 GeV/c) to high (50 GeV/c) transverse momenta
charged jets were defined using only charged particles
of $p_T > 0.5$ GeV/c in a cone clustering algorithm.
Evidence of charged jets has been observed in minimum
bias data already around 1--2 GeV/c, and their properties 
join smoothly these for
jets observed in regular jet-triggered data ($>20$ GeV/c).
This charged jet technique was also employed
for investigating properties of the UE and
of minimum bias events. The analysis required a 
reconstructed charged jet, and the phase space was 
divided into angular regions ``toward", ``away" and
``transverse" to the jet, the latter being of primary
interest here.
While the two former regions were fairly well described by \pythia\ and
\herwig, none of the MCs examined described correctly all the properties
of the UE (i.e. the ``transverse'' region).
The influence of the UE on jet energies was also studied in the high-$E_T$
calorimeter-jet data by defining two cones at the same $\eta$, but
at $\pm 90^{\circ}$ in azimuth from the leading jet. 
The lower energy ($min$) cone was used to estimate the contribution
from UE, while the higher ($max$) cone also receives contributions 
from perturbative radiation (thus measured by the $max-min$ difference).
The study concluded that the UE in hard scattering events is more active
than in soft collisions. In general, these results will be useful
for tuning the MC generators.

\section{Heavy Flavor Production and Decay}

With the observation of large discrepancies between 
data and theory
Heavy Flavor Production is currently one of the most 
interesting topics of High Energy Physics.
At this workshop new results concerning charm, beauty
and charmonium production at HERA and LEP were reported.
>From the experimental point of view, the improvement
in background reduction for the measurement of charm 
mesons, as obtained by the application of a decay length
cut using the H1 Central Silicon Detector, is particularly
impressive, see the talk\cite{wagner} by \textbf{Jeannine Wagner}. 
On the theoretical side, applications of the saturation 
model and the $k_T$ factorization model to heavy flavor
production, as presented during this workshop, are particularly
noteworthy.

\subsection{Charm production}

In general, the agreement between theory and 
measurement of open charm production is satisfactory.
NLO calculations of charm production
in deep inelastic scattering are able to reproduce the
measured rates of $D^*$ production \cite{h1-z-d1}.
In photoproduction the predictions\cite{frixione1} to 
fixed-order plus next-to-leading logarithms agree with 
measurements of H1 at two different $\gamma p$ center 
of mass energies, but lie a factor 1.5 to 2 below measurements
of the ZEUS collaboration.
As of now, there are no published
measurements of open charm production cross sections
available from the Tevatron collider.
Given the large differences between measurement and calculation
for beauty and charmonium production at the Tevatron,
measurements of charm production cross sections in $p\overline{p}$ 
are eagerly awaited.

With the large sample of charm mesons collected at HERA it is
possible to extract charm fragmentation factors and parameters
with a precision comparable to $e^+e^-$ experiments.
\textbf{Jeannine Wagner} presented\cite{wagner} new measurements by H1 of
the fragmentation factors $f(c\rightarrow D)$ for $D^+$, $D^0$, 
$D_s$ and $D^*$ mesons.
The results are in excellent agreement with the values obtained
elsewhere and can be used to extract various charm
fragmentation parameters, such as $P_v$ (the fraction of vector mesons), 
$R_{u/d}$ (the ratio of u and d quarks) and $\gamma_s$ (the strangeness
suppression factor).
For instance, assuming isospin invariance the fraction of vector
mesons is determined to be
$P_v = 0.613\pm 0.061 (stat) \pm 0.033 (syst) \pm 0.008 (theo)$.
\textbf{Sanjay Padhi} reported on a similar determination based on ZEUS
data, quoting $P_v = 0.546 \pm 0.045 (stat) \pm 0.028 (syst)$ \cite{padhi}.
Both results are in excellent
agreement with the world average of $P_v = 0.601 \pm 0.032$.

Various processes contribute to the photoproduction of
charm quarks: in direct photoproduction the dominant process is 
$\gamma - g$ fusion with a quark
exchange, in resolved photoproduction charm excitation with a gluon 
exchange is expected to dominate.
Due to the different spins of quarks and gluons, the angular
distribution of the two processes differ, leading to an excess of
events in the $\gamma$-direction for resolved events.
A study\cite{padhi} of the angular distribution of dijet events with an identified
$D^*$ meson based on ZEUS data was presented by \textbf{Sanjay Padhi}.
The angular distributions for the direct and resolved photon enriched
samples are observed to be significantly different. 
A significant excess of events is observed in the $\gamma$ direction 
for the sample enriched in resolved photoproduction events.
This may be interpreted as the first convincing evidence for charm
excitation in the photon.

\textbf{Stephen Robins} presented new measurements of $D^*$ production
cross sections in deep inelastic scattering by the ZEUS collaboration.
The results\cite{robins} are based on the large data sets collected in
the 1998-2000 running periods, corresponding to 82 $pb^{-1}$.
Overall the data exhibit the features expected from NLO
pQCD predictions based on recent parton density functions, such as CTEQ5F3.
However, the data for electron- and positron-proton scattering show
a statistically somewhat significant difference: the ratio of $e^-$ and
$e^+$ events increases with momentum transferred, from about 1.25 at
$Q^2 = 30$ GeV$^2$ to 2.25 at $Q^2 = 450$ GeV$^2$.
Detailed systematic checks have been performed to exclude differences
in the acceptance of $e^-p$ and $e^+p$ collision events.
No plausible physics reason for the observed difference has yet been proposed.

A wealth of results on charm production in $\gamma\gamma$ interaction 
is emerging from the LEPII data.
\textbf{Armin B\"ohrer} reported on $D^*$ rates versus
pseudorapidity, $\eta$, transverse momentum of the $D^*$, $p_T^{D^*}$,
momentum fraction of the photon partaking in the
hard interaction, $x_\gamma$, visible $\gamma\gamma$ center-of-mass 
energy, $W_{vis}$ and $e^+e^-$ center-of-mass energy, $\sqrt{s}$
\cite{bohrer}.
The agreement with NLO pQCD predictions is very good, apart
from the differential cross section versus $x_\gamma$, as measured
by OPAL, which shows an excess at a low value of $x_\gamma =0.03$.
The excess might be explained by additional hadron like contributions
not included in standard parton density functions of the photon.

Motivated by the success of the Saturation Model in describing the
proton structure function in the transition region between photoproduction
and deep inelastic scattering and inclusive diffractive scattering in 
deep inelastic scattering, \textbf{Antoni Szczurek} extended the
formalism to include heavy quark production in $\gamma$ - nucleon and
$\gamma\gamma$ scattering\cite{szczurek}.
The saturation model describes the scattering of photons with the
nucleon as convolution of a perturbatively calculable transverse
photon wave function with a dipole-nucleon cross section parametrized
in fits to proton structure function data. 
In its extension to describe heavy flavor production, particular care was 
devoted to the effects of the kinematical threshold
and the phase-space limitations in the region of large $x_\gamma$.
The calculations are able to describe measurements at low $W_{\gamma p}$,
but lay significantly below the measurements from HERA with $W_{\gamma p}
 > 100$ 
GeV.
The situation for charm production in $\gamma\gamma$ collisions is somewhat
better.
The calculations are able to reproduce the shape of the rising cross section 
with $W_{\gamma \gamma}$, however the absolute normalization appears to be
about 30\% below the data.

Traditionally, inclusive DIS is analyzed within the framework
of the DGLAP evolution 
scheme\cite{Gribov:1972ri,Lipatov:1975qm,Altarelli:1977zs,Dokshitzer:1977sg}. 
In a different formalism, the $k_T$ factorization scheme,
terms proportional to $\alpha_s \log(\mu/\Lambda)^2 \log(1/x))^n$
are resummed to all orders and give rise to a
so-called unintegrated gluon momentum distribution, $A(x,k_T^2,\mu^2)$.
The latter is dependent on $x$, the momentum fraction of the gluon 
in the proton, the probing scale $\mu$ and the transverse momentum
$k_T$ of the gluon.
\textbf{Nikolaj Zotov} presented a calculation\cite{zotov} of charm production
cross sections based on three different choices of unintegrated
gluon momentum distributions: a) by J Bl\"umlein\cite{jb}, based on the leading
order solution of the BFKL\cite{Kuraev:1977fs,Balitsky:1978ic} equation;
b) by J Jung and G Salam\cite{js}, based on a numerical
solution to the CCFM\cite{ccfm} equation;
and c) by M Kimber {\em et al.}\cite{kmr}, based on a 
combination of the BFKL and DGLAP equations.

Despite large differences among the 
three choices of unintegrated gluon distributions, 
the corresponding predictions for the charm 
production cross sections in both photoproduction and DIS are remarkably
similar and agree well with measurements by the HERA collider experiments.

\subsection{Beauty production}

Production of beauty quarks is an area where the predictions of pQCD
are expected to be accurate and reliable.
Beauty production cross sections have been measured in $\gamma \gamma$, 
$\gamma p$, $\gamma^* p$, and $p \overline{p}$ collisions. 
Strikingly, in all these environments the experimental results lie factors of 
3 - 5\cite{frix2,cdf} above next-to-leading order pQCD predictions. 

In order to understand the reasons for the failure of the calculations, the
experiments are devoting large efforts to cross check their measurements,
using e.g. other detection methods, and/or to extend the kinematical range of their
measurements. 
In this context \textbf{Monica Turcato} presented a recent
analysis\cite{turcato} of photoproduction events by the ZEUS collaboration. 
The events containing b-quarks are identified through the measurement of
the relative 
momentum of the decay $\mu$ with respect to the axis of its associated jet. 
The analysis makes use of the forward muon chambers, thus extending the 
angular coverage to pseudorapidities of 2.3. 
The results compare well to leading-order MC, such as \pythia. 
Measurements of the differential cross sections and comparison to NLO pQCD
calculations are forthcoming.

Another way to obtain a handle on the b-quark production cross section
was presented by \textbf{Jeannine Wagner}.
The analysis\cite{wagner} 
investigates the angular correlation of $D^*$ mesons and $\mu$'s in events containing
both particles.
The event sample is subdivided in four subsamples with either same/opposite charge
and same/opposite side $D^*$'s and $\mu$'s.
The sample of same side and same charge particles is almost entirely originating
from B meson decays, with only a small contamination of charm quarks and `fake' $\mu$'s.
The extracted visible b quark production 
cross section $\sigma_{vis}(ep \rightarrow e'D^* \mu X) = (380 
\pm 120 \pm 130)$ pb is significantly above the predictions of 106 pb by the \AROMA\
leading-order MC.

The situation is very similar in the $\gamma\gamma$ environment\cite{bohrer}.
\textbf{Arnim B\"ohrer} presented recent measurement by L3 and OPAL of the
inclusive beauty cross section.
The measurements are based on an analysis of the relative momentum of either an
electron or muon with respect to the axis of an associated jet.
The measurements cluster around 13 pb and are about a factor three above the
NLO pQCD prediction, a difference which corresponds to four standard deviations.

Given the many significant discrepancies between experiment and theory in beauty production, 
one would expect a large theoretical effort aiming at resolving these differences.
Disappointingly, the workshop received only one contribution in which 
\textbf{Antoni Szczurek}
applied\cite{szczurek} the saturation model (see above) to beauty photoproduction. 
Similarly to the collinear approach, his results are about a factor 2-3 below the
recent HERA measurements at $W_{\gamma p} \sim$ 190 GeV.
To help resolve this discrepancy, measurements at different $W_{\gamma p}$ 
would be particularly useful.

\subsection{Charmonium production}

\textbf{Jungil Lee} presented a comprehensive overview of the status of theoretical calculations
of $J/\psi$ - meson production.
He reviewed\cite{lee} the physics of the non-relativistic QCD factorization approach 
and the contributions of color-singlet (CS) and color-octet (CO) processes.
Due to their different dependences on the transverse momentum $p_T$ of the $J/\psi$ meson the
sum of these processes is able to describe the differential cross section versus $p_T$, 
as measured by CDF at the Tevatron.
The strength of the CO contribution is described by universal matrix elements which can
not (yet) be calculated theoretically, but have to be determined by experiment.

\textbf{Katja Kr\"uger} presented new results\cite{kruger} from the H1 collaboration
concerning inelastic $J/\psi$ photo- and electroproduction.
The measurements span a large region of $z$, the inelasticity of the $J/\psi$ mesons, defined as the
ratio of the energies of the $J/\psi$ meson and the photon in the proton rest frame.
The results in photoproduction are well described by NLO pQCD calculation 
and the CS model.
The importance of NLO corrections is particularly evident in the description of the 
$p_{T,\psi}^2$, where a leading order calculation fails to describe the shape of the cross
section.
Unfortunately, the NLO calculations suffer from a large uncertainty due to the wide range
of allowable masses of the charm quark.
When compared to leading order calculations, the differential cross sections are compatible 
with small values of the CO matrix element, but do not appear to be in contradiction with
the values determined from the Tevatron data.
Clearly, a rigorous calculation to NLO including both CS and CO contributions
is needed for both the HERA and Tevatron environment before firm conclusions about the
universality of the matrix element can be drawn.

The measurement of the helicity distribution of $J/\psi$ mesons versus $z$ in photoproduction 
shows a different trend than LO calculations including CS or CS+CO contributions.
The data tend to go from a full transverse polarization at low $z$ to no polarization at high $z$,
whereas the calculations seem to do the inverse.
However, one has to keep in mind that
the calculations are to leading order and do not resum soft gluon emissions.
This may be an important shortcoming of the calculations, 
since each emitted gluon carries away one unit of spin.

\textbf{Arnim B\"ohrer} reported on recent calculations\cite{klasen} 
of the differential cross section versus
$p_T^2$ for inelastic $J/\psi$ production in the $\gamma\gamma$ environment\cite{bohrer}.
The calculations use the matrix elements determined by Tevatron data to estimate the
CO contribution and is in good agreement with the data.
However, the data are not precise enough to unambiguously 
establish the presence of CO contributions.
Furthermore, the calculations are to LO only and large NLO corrections may 
affect the present results.

\section{Dimuon photoproduction}

H1 measured\cite{leissner} di-muon photoproduction in search of deviations from Standard Model
expectations.
\textbf{Boris Leissner} presented the measured
differential cross section versus the mass of the dimuon system.
No obvious deviation from expectations, calculated with the \GRAPE\ MC, are observed.
The visible cross section, both inclusive and for the inelastic subsample, are in 
excellent agreement with the predictions.

\section{Instanton searches at HERA}

Instanton induced events are
expected\cite{koblitz,Moch:1996bs,Ringwald:1998ek} to leave a distinct
signature: high multiplicity of tracks in the forward region, flavor
symmetry, large muon content etc.  \textbf{Birger Koblitz} applied a
multi-variant discrimination method to obtain limits on instanton
induced processes.  The analysis is performed in a corner of phase
space where backgrounds from non-instanton induced events are
minimized.  Based on H1 data, exclusion limits of 55 to 80 pb are
obtained if backgrounds from non-instanton induced processes, as
predicted by standard LO MCs, are subtracted.  Under the assumption
that all selected events are instanton induced, the limit is degraded
to 255 pb.

\section{\boldmath Small-$x$ final states}
\label{sec:small-x}

As measurements from HERA of small-$x$ final states have accumulated,
it has become very clear that we do not understand the details of
partonic evolution in this region of phase space. Even though NLO
DGLAP~\cite{Gribov:1972ri,Lipatov:1975qm,Altarelli:1977zs,Dokshitzer:1977sg}
evolution is able to reproduce the increase of $F_2$ with decreasing
$x$, it is clear from final-state data that parton evolution is more
complicated. The alternative,
BFKL~\cite{Kuraev:1977fs,Balitsky:1978ic} evolution, has turned out to
suffer from huge NLO corrections \cite{Fadin:1998py} and there does
not seem to be any completely satisfactory model available for
small-$x$ final states today (see e.g.\ \cite{Anderson:2002cf} for a
recent review).

One problem with the BFKL formalism is that it does not conserve
energy and momentum. Although this may formally not be a problem at
asymptotically large energies, in the real world it may make a very
large difference. This was clearly demonstrated in the presentation by
\textbf{Jeppe Andersen} \cite{DIS02Andersen}. By implementing BFKL
evolution in an event generator where energy and momentum is conserved
exactly in each vertex, he showed that the predicted BFKL enhancement
of di-jet cross sections with the rapidity difference between the
jets, actually transforms into a suppression. The reason is that if
you take into account the energy needed to emit the gluons
responsible for the increase of the partonic cross section, you will
find that the parton densities are probed at correspondingly higher
$x$-values. And since the gluon density decreases steeply with $x$,
the total effect is a suppression. Andersen also suggested
measurements which involved incoming quarks, the densities of which do
not decrease as much with $x$.

When going to next-to-leading order in the evolution
\cite{Fadin:1998py}, some energy-momentum conservation effects are
taken into account. But to get reliable predictions it is necessary to
also calculate the so-called impact factors (or off-shell matrix
elements) to next-to-leading order. This has previously been done for
the $\gamma^*$ impact factors \cite{Bartels:2000gt}, but to describe
di-jets at large rapidity differences one needs to calculate the jet
impact factors to NLO. \textbf{Gian Paulo Vacca} presented a first
calculation for this \cite{DIS02Vacca}. For the $\gamma^*$ impact
factors one must treat carefully the cancellation between real and
virtual diagrams in the ladder and the corresponding ones from the
scattered $q\bar{q}$ system. For the jet impact factors there is the
additional complication of also considering real and virtual diagrams
in the parton densities. At the time of the presentation, the
calculations were not quite ready and no predictions were presented,
but since then the calculation has been completed
\cite{Bartels:2002yj}.

Further theoretical progress in understanding small-$x$ evolution was
presented in the \textit{Structure Functions} parallel sessions
\cite{DIS02STRF}. Also some new measurements were presented. Relevant
for this summary was the report from \textbf{Lidia Goerlich} on forward
pion production at HERA \cite{DIS02Goerlich}. The advantage of looking
at $\pi^0$ production rather than at jets, is that it is possible to
go further forward and hence be more sensitive to small-$x$ dynamics.
The downside is, of course, that one needs to understand the
fragmentation of partons into pions. The presented measurement is a
update of a previous measurement \cite{Adloff:1999zx} with increased
statistics and a more detailed study of the event structure. The
result is that, similar to forward jet measurements, it is impossible
to describe the rate and distribution of forward $\pi^0$ with standard
DGLAP based Event Generators, while adding a resolved virtual photon
component does quite well. To some surprise, the \cascade program
\cite{Jung:2001hx}, which is able to describe forward jets, is not
able to describe forward pions. The problem may be that the program
does not include quarks in the evolution. The investigation into the
distribution of particle flow in events with a forward pion was also
presented, but no additional discrimination power between models was
obtained.

\section{Hadronization}
\label{sec:hadronization}

One problem with small-$x$ evolution is that it balances on the border
between perturbative and non-perturbative QCD, and our understanding
of the latter is very poor. For the hadronization process we have a
couple of models, string \cite{Andersson:1983ia} and cluster
\cite{Marchesini:1988cf} fragmentation, which are able to describe
most features of the hadronic final-states in \ee, but it is not
entirely clear if these models can be used without modification in
collisions involving incoming hadrons.

\textbf{Dimitry Ozerov} presented an investigation of the
fragmentation parameters in the \pythia program
\cite{Sjostrand:2000wi} and found that he needed a smaller value of
the parameter controlling the di-quark production in the string
braking to reproduce the production of (anti-) protons in
photoproduction events at HERA as compared to the value fitted to LEP
data \cite{DIS02Ozerov}. It was noted, however, that he had only been
looking at minimum bias events, which are known to be notoriously
difficult to model. Ozerov also noted an interesting scaling property
when looking at the $m_\perp=m+p_\perp$ distribution scaled with the
number of spin states, which seemed to be independent of particle
species.

Looking at high-$E_T$
jets and DIS events, where there is a hard
scale present, the fragmentation of jets are assumed to be more
similar to the \ee case. Indeed, contrary to the minimum bias case
above, there does not seem to be any need for a very different
di-quark suppression in string fragmentation. At least not when
comparing the production of kaons and lambdas in photo-production as
presented by \textbf{Stewart Boogert} \cite{DIS02Boogert}. On the
other hand, he finds a need for stronger strangeness suppression in
the string fragmentation (again compared to LEP data) when looking at
$\phi(1020)$ production in DIS, although there were some questions
raised about the normalization uncertainties. Since the $\phi(1020)$
is a pure \qs\sbar\ state, Boogert also showed that he can use such
particles with very large fraction of the virtual photon momentum to
directly probe the strange sea distribution.

\textbf{Erika Garutti} presented an interesting investigation of the
formation time of hadrons \cite{DIS02Garutti}. Looking at pions, kaons
and protons produced in DIS on heavy nuclei at HERMES, and comparing
the attenuation in different nuclei as a function of their speed, she
showed that protons seem to have a longer formation time than pions
and kaons. Although there may still be some model dependency in the
conclusions, these findings will be important for e.g.\ finding
evidence for quark-gluon plasma.

Another result from HERMES was presented by \textbf{Volodymyr
  Ulesh\-chenko} \cite{DIS02Uleshchenko}. Looking at particles
carrying a large fraction of the virtual photon momentum, one would
expect that the ratio of charged to uncharged pions would be constant,
if one only considered the na{\"\i}ve parton model. This is, however,
not what has been found at HERMES. Ulesh\-chenko showed that the
discrepancy can be explained by adding a diffractively produced
$\rho^0$ decaying into $\pi^+\pi^-$. He also noted that this may
influence previously published results on the flavour asymmetry of the
light sea quarks \cite{Ackerstaff:1998sr}.

Another way of getting insight into the particle production mechanism
is to look at particle correlations. It is well known that the
production of identical bosons should be enhanced if the particles are
close in phase space. This so-called Bose-Einstein correlation depends
on the size of the production region and can thus tell us about how
particles are produced. \textbf{Edward Sarkisyan-Grinbaum} reviewed
some results on multi-particle correlations from LEP
\cite{DIS02Sarkisyan}. He showed that the size of the production
region, as obtained from Bose-Einstein correlations, seems to depend
on the mass of the produced hadron, which casts some doubt on whether
we really understand what is measured. He also showed results on
genuine 3- and 4-particle correlations which are consistent with an
incoherent source of particles. An interesting fact is that in fully
hadronic W$^+$W$^-$ events, where we expect the hadronization region
from the two W's to overlap, there seems to be no \textit{cross-talk}
whatsoever between the W's. Note that even if particles are produced
completely independent from the W's we still expect Bose-Einstein
correlations, but none has been found.

\textbf{Malcolm Derrick} reported on Bose-Einstein correlations at
HERA \cite{DIS02Derrick}. Here one could imagine a decreasing size of
the production region with increasing $Q^2$, but no such dependence
was found. In photo-production there does, however, seem to be a
larger size of the production region.

\section{Resummation and parton showers}
\label{sec:resumm-part-show}

By looking at infrared safe event shapes it is possible to avoid event
generator based hadronization models altogether and still be able to
obtain reliable QCD predictions. This is done by careful resummation of
perturbative diagrams, together with a simple perturbatively inspired
model to take \textit{power corrections} from hadronization effects
into account. \textbf{Gavin Salam} gave a brief overview of the
results for wide range of event shapes measured at LEP, where it is
possible to obtain a consistent fit to the two parameters, \as and
$\alpha_0$ \cite{DIS02Salam}. The main part of his presentation was
aimed towards DIS, where one na{\"\i}vely would expect the current
hemisphere of the Breit frame to look like half an \ee event. There
are, however large spill-over effects from the target hemisphere, the
phase space of which becomes very large at small $x$, and Salam
presented a way to overcome these problems. He also presented a new
program called \dispatch with a common interface to the \disent
\cite{Catani:1997vz} and \disaster \cite{Graudenz:1997gv} NLO
programs.

\textbf{Giulia Zanderighi} also presented results on event shapes in
DIS, but while most such investigations so far have dealt with two-jet
shapes, she has been looking at the three-jet related $K\sub{out}$
shape \cite{DIS02Zanderighi}. This measures the momenta out of the
plane defined by the $\gamma^*-$p and thrust-major axis. The
calculation involves resummation of large single and double logarithms
of $K\sub{out}/Q$ which depend on the geometry of two outgoing jets,
the colour structure, the parton densities and the experimental
acceptance in the forward region. She also presented an effort to
produce a MC program to semi-automatically calculate resummed
power-corrected event shape predictions for any hard interaction in
\ee, DIS and hadronic collisions.

Another three-jet related event shape in DIS was discussed by
\textbf{Andrea Banfi} \cite{DIS02Banfi}. He presented a calculation of
the azimuthal correlations in DIS, which is similar to the
energy-energy correlations in \ee, but uses transverse momenta in
the Breit frame rather than energies. One peculiarity of this
event shape is that it does not go to zero with the azimuthal angle,
which results in a non-integer power correction. This event shape has
not yet been measured at HERA, but now that a prediction exists, such
a measurement is surely called for.

There were two presentation about new algorithms for combining fixed
order matrix elements with parton showers. Fixed order matrix can be
used to describe the production of a handful well separated partons,
but cannot be used to describe the soft and collinear inter- and
intra-jet partons. For the parton shower models, the situation is
reversed, which is why algorithms combining the two are very useful.
Such algorithms are, however, not trivially constructed.

The first of these presentations, by \textbf{Frank Krauss}
\cite{DIS02Krauss}, described an algorithm based on the angular
ordered parton shower in \apacic\cite{Kuhn:2000dk}. Using a 2-,
3-,\ldots,$n$-jet matrix element generator for \ee, the events
obtained are reweighted with the relevant Sudakov form factors from
the parton shower. To do this he needs to use the $k_\perp$-algorithm
\cite{kTalgorithm} to reconstruct the emission scales. After this, a
special \textit{vetoed} version of the parton shower is added before
hadronization can be performed. In this way the dependence on the
regularization scale in the matrix element is canceled to
next-to-leading logarithmic accuracy \cite{Catani:2001cc}. Krauss also
presented a strategy for extending this algorithm to hadronic
collisions \cite{Krauss:2002up}.

\textbf{Leif Lönnblad} presented a similar algorithm
\cite{DIS02Lonnblad,Lonnblad:2001iq} also based on the ideas in
\cite{Catani:2001cc}.  The main difference was that the Colour Dipole
Model as implemented in \ariadne \cite{Lonnblad:1992tz} was used for
the partonic cascade rather than the \apacic parton shower. Also a
modified version of the \diclus algorithm \cite{Lonnblad:1993qd} was
used to reconstruct, not only emission scales, but complete dipole
cascade histories of the matrix element generated states. The Sudakov
form factors are then generated from within the dipole cascade with a
special veto algorithm.

\providecommand{\href}[2]{#2}\begingroup\raggedright\endgroup

\end{document}